\documentstyle[12pt,axodraw,epsf,epsfig]{article}

\begin{document}

\begin{titlepage}

\begin{center}

\hfill ICRR-Report-519-2005-2\\
\hfill \today

\vspace*{1cm}

{\large 
Hadronic EDM Constraints on Orbifold GUTs
}
\vspace{1cm}

{\bf Junji Hisano},  
{\bf Mitsuru Kakizaki}
and
{\bf Minoru Nagai}
\vskip 0.15in
{\it
{ICRR, University of Tokyo, Kashiwa 277-8582, Japan }
}
\vskip 0.5in

\abstract{
We point out that the null results of the
hadronic electric dipole moment (EDM) searches
constrain orbifold grand unified theories (GUTs), 
where the GUT symmetry and supersymmetry (SUSY) are both
broken by boundary conditions in extra dimensions 
and it leads to rich fermion and sfermion flavor structures.
A marginal chromoelectric dipole moment (CEDM) of the
up quark is induced by the misalignment between the CP violating
left- and right-handed up-type squark mixings,
in contrast to the conventional four-dimensional SUSY GUTs.
The up quark CEDM constraint is found to be as strong
as those from charged lepton flavor violation (LFV) searches.
The interplay between future EDM and LFV experiments 
will probe the structures of the GUTs and 
the SUSY breaking mediation mechanism.
}

\end{center}
\end{titlepage}
\setcounter{footnote}{0}

After the discovery of successful agreement of the three extrapolated 
gauge couplings at a higher energy scale, 
supersymmetric grand unified theories (SUSY GUTs) have been considered
as ones of the most predominant candidates for theories
beyond the standard model (SM).
The ratio of the bottom quark to tau lepton masses 
supports the idea of GUTs which also predict the Yukawa coupling
unification.
There have been extensive researches
to detect further implications for SUSY GUTs.

Crucial hints for SUSY GUTs are also obtained
in flavor and CP violation 
originated from the SUSY breaking scalar mass terms.
Because SUSY GUTs also provide some relations between squark and slepton
mass matrices \cite{Hisano:2003bd,Ciuchini:2003rg},
we can probe structure of SUSY GUTs
by combining results of low energy experiments,
such as $K$, $B$ physics, lepton flavor violating (LFV) decays and 
electric dipole moment (EDM) searches.

Recently it was pointed out that
null results of hadronic EDM experiments severely constrain 
CP-violating squark flavor mixings in SUSY models 
\cite{Hisano:2003iw,Hisano:2004tf}.
When off-diagonal elements in squark mass matrices have sizable 
imaginary parts,
unacceptably large chromo-electric dipole moments (CEDMs) would be
induced and conflict with hadronic EDM data.
The flavor conserving moments, which are induced by loop diagrams,
depend on the flavor violating mass parameters via the internal lines,
and they are enhanced by heavier quark masses due to
a chirality flip nature of the dipole moment operators.
In particular the bound on down-type squark mixings
between the second and
third generations plays a crucial role in discussing 
the SUSY contribution to the CP asymmetry in the $B$ meson system.
If a sizable deviation from the SM prediction of
the CP asymmetry in $B^0_d \to \phi K_S$ \cite{BphiK}
is attributed to
the right-handed down-type squark mixing,
which is radiatively generated by
the right-handed neutrino Yukawa coupling in SUSY GUTs 
\cite{Moroi:2000tk},
the CEDM of the strange quark would exceed the experimental bound.
We emphasize that 
the hadronic EDM constraints are important criteria for distinguishing
various scenarios and for building models 
\cite{Hisano:2003iw,Hisano:2004tf,Dimopoulos:1994gj,Khriplovich:1996gk,Romanino:1996cn,Hisano:2004pw}.

In constructing realistic SUSY GUT models,
we are confronted with many difficulties.
Baryon number violating dimension five operators induced by 
color triplet Higgs multiplet exchange have to be suppressed
in order to stabilize proton adequately \cite{proton_decay}.
The mass ratios of the down-type quarks to charged leptons
in the first two generations are quite different from the counterpart
of the third generation.
In addition, 
a vast mass hierarchy between the light Higgs doublets and
their heavy color triplet partners is required without invoking
a fine-tuning of parameters.

There are many attempts to solve the problems inherent in SUSY GUTs.
Approaches employing extended
representations and/or groups are well investigated \cite{dim5sup}.
Meanwhile quite different ideas utilizing spatial extra dimensions
were proposed and developed in recent years 
\cite{Kawamura:2000ev,Altarelli:2001qj,Hall:2001pg,Hebecker:2001wq,Hebecker:2001jb,Asaka:2001eh,Hall:2001rz,Hall:2001xb,Hall:2002ci}.
The GUT gauge symmetry manifest in higher spacetime dimensions
is broken to the SM ones by the boundary conditions 
on a compactified orbifold.
The higher dimensional SUSY is also reduced by the compactification.
In this construction the mass hierarchy between doublet and triplet
Higgs fields is realized without any redundant Higgs multiplets.
Furthermore the dangerous dimension five proton decay and
the wrong prediction of Yukawa couplings in the first two generations
are avoided by appropriate choices of matter field configuration,
while the gauge coupling unification 
at the cutoff scale $M \sim 10^{17}$ GeV is maintained
\cite{Hall:2001pg,Hall:2001xb,Hall:2002ci}.

In this letter we discuss the hadronic EDMs 
in the orbifold GUT models which explain 
flavor structures of the fermions and sfermions
by the geometrical nature of higher dimensional spaces.
For definiteness we will work in a framework of SU(5) unification 
proposed by Hall and Nomura \cite{Hall:2002ci}.
In this setup 
the matter ten-plets of the first two generations reside in the bulk,
while other matter fields are located on the SU(5) preserving brane.
The soft SUSY breaking mass terms for bulk fields 
are attributed to the Scherk--Schwarz mechanism
\cite{Scherk-Schwarz,Barbieri:2001yz}
with no soft masses for brane fields, 
resulting in non-universal mass spectrum for left-handed squarks,
up-type squarks and right-handed sleptons.
In generic the CP violating phases of the off-diagonal
elements in the left-handed up-type quark mass matrix 
do not coincide with
the right-handed counterparts, and they lead to the up quark CEDM
enhanced by the top quark mass. 
We find that the predicted up quark
CEDM is marginal to the current experimental limit. 
In this model observable
lepton flavor violating processes are also predicted due to 
the off-diagonal
terms in the right-handed slepton mass matrix. 
The constraint from the up quark CEDM on the orbifold GUT model 
is as strong as that from $\mu \rightarrow e \gamma$.

While the up quark CEDM could gives a severe constraint on the SUSY
models potentially, the CEDM is suppressed under the universal scalar
mass hypothesis. 
The orbifold GUT models, which we discuss in this paper,
are the exceptions.

The proposed deuteron EDM search, 
which is expected to reach to the $10^{-27}\ e \ \mbox{cm}$ level,
is very promising \cite{Semertzidis:2003iq}.
The limit on the T-odd nuclear force will be lowered
by two orders of magnitude over the current limit.
Thus the structures of sfermion mass matrices will be well understood
even if calculation of the deuteron EDM involves
theoretical uncertainties arising from nuclear dynamics.

First, we recall the CEDMs of the light quarks and
their experimental upper limits constrained by hadronic EDM experiments.
There are several contributions to observable hadronic EDMs.
The CP violating effects stem from the CEDMs $d^C_q$ and the EDMs $d_q$ 
of the light quarks $q (=u,d,s)$, and the QCD $\theta$ parameter
up to the dimension five level.
Using the SU(3) chiral Lagrangian technique,
the quark CEDM contributions 
to the EDMs of the neutron and $^{199}$Hg atom are evaluated as
\cite{Hisano:2004tf}
\begin{eqnarray}
  d_n & = & 
  - 1.6 \times e (d_u^C + 0.81 \times d_d^C + 0.16 \times d_s^C),
  \nonumber \\
  d_{\rm Hg} & = &
  - 8.7 \times 10^{-3} \times 
  e (d_u^C - d_d^C + 0.005 \times d_s^C),
\end{eqnarray}
when the Peccei--Quinn (PQ) mechanism works.
Here $d_n$ is
induced by the charged meson loop processes,
and $d_{\rm Hg}$ arises from the nuclear force generated by
meson exchanges.
The quark CEDM contributions to hadronic EDMs depend on
whether the PQ symmetry works to relax the strict bound on
the $\theta$ term or not \cite{Bigi:1991rh}, 
although the numerical difference is small.
The effects induced by the usual EDMs $d_q$ of the light quarks to $d_n$
are found to be comparable to those from the quark CEDMs 
in several approaches \cite{Pospelov:2005pr}.
Meanwhile $d_{\rm Hg}$ is almost independent of the quark EDM operators
due to the Schiff screening.
Since we pay the attentions to the order of magnitudes of the hadronic
EDM sensitivities, 
in this letter we assume the PQ symmetry and discard the
quark EDM contributions for simplicity.
The current upper bounds on the EDMs of the neutron \cite{Harris:1999jx}
and the $^{199}$Hg atom \cite{Romalis:2000mg} are 
\begin{eqnarray}
  |d_n| < 6.3 \times 10^{-26} \ e\ \mbox{cm}, \quad
  |d_{\rm Hg}| < 1.9 \times 10^{-28} \ e\ \mbox{cm}
\end{eqnarray}
at $90\ \%$ confidence level, respectively.
In the absence of accidental cancellation among various contributions
we obtain
\begin{eqnarray}
  e|d^C_u| & < & 3.9 (2.2) \times 10^{-26}\ e\ \mbox{cm}, \nonumber \\
  e|d^C_d| & < & 4.8 (2.2) \times 10^{-26}\ e\ \mbox{cm}, \nonumber \\
  e|d^C_s| & < & 2.4 (44) \times 10^{-25}\ e\ \mbox{cm}  
\end{eqnarray}
from the EDM constraints of the neutron ($^{199}$Hg atom).

In discussing flavor violation in SUSY models,
it is convenient to introduce the following mass insertion parameters
\cite{mass_insertion}:
\begin{eqnarray}
  (\delta^f_{ij})_{L/R} \equiv
  \frac{(m_{{\tilde{f}_{L/R}}}^2)_{ij}}{(\overline{m}_{{\tilde{f}}}^2)},
  \quad 
  ( \delta_{i}^{d})_{LR} \equiv
  \frac{m_{d_i}(A_i^{(d)} - \mu\tan\beta)}{\overline{m}^2_{\tilde{d}}},
  \quad
  ( \delta_{i}^{u})_{LR} \equiv
  \frac{m_{u_i}(A_i^{(u)} -\mu\cot\beta)}{\overline{m}^2_{\tilde{u}}},
\end{eqnarray}
where $i,j=1,2,3$ denote the flavor indeces.
Here these quantities are defined 
in the so-called super-CKM basis,
where the corresponding quark mass matrices are diagonal
and flavor violating effects are encoded into squark mass matrices.

\begin{figure}[t]
  \begin{center} 
    \begin{picture}(455,140)(30,-20)
      \SetOffset(100,0)
      
      \ArrowArcn(135,25)(75,180,90)
      \ArrowArc(135,25)(75,0,90)
      \Vertex(135,100){3}
      \Text(135,120)[]{$\tilde{g}$}
      
      \ArrowLine(60,25)(30,25)
      \DashArrowLine(90,25)(60,25){3}             \Vertex(90,25){3}
      \DashArrowLine(135,25)(90,25){3}             \Vertex(135,25){3}
      \DashArrowLine(135,25)(180,25){3}       \Vertex(185,25){3}  
      \DashArrowLine(180,25)(210,25){3}        
      \ArrowLine(210,25)(240,25)
      
      \Text(45,15)[]{$q_{Li}$}
      \Text(70,15)[]{$\tilde{q}_{Li}$}
      \Text(115,15)[]{$\tilde{q}_{Lj}$}
      \Text(155,15)[]{$\tilde{q}_{Rj}$}
      \Text(200,15)[]{$\tilde{q}_{Ri}$}
      \Text(225,15)[]{$q_{Ri}$}
      
      \Photon(172,110)(195,135){2}{5}
      \Text(203,145)[]{$g$}
    \end{picture} 
    \caption{{\footnotesize
        Diagram contributing to the CEDM of the 
        quark when both the left-handed and right-handed squarks have 
        CP violating flavor mixings.
        Notice that these diagrams are enhanced 
        by the heavier quark mass 
        due to the chirality flip in the internal line.
    \label{fig:CEDM}
    }}
  \end{center}
\end{figure}
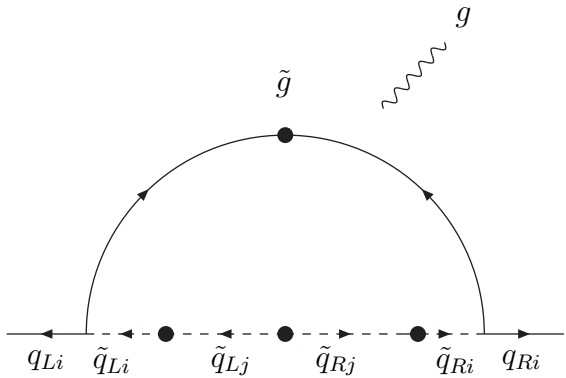

When both the left-handed and right-handed squarks have
off-diagonal components in their mass matrices, 
the CEDMs of the light quarks are significantly enhanced
by the heavier quark mass.
From Figure \ref{fig:CEDM} we can estimate the quark CEDM
originated from the squark flavor mixings as
\begin{eqnarray}
  d^C_{q_i} \sim 
  \frac{\alpha_s}{4\pi}\frac{m_{\tilde g}}{\overline{m}^2_{\tilde q}}
  {\rm Im}
  \left[(\delta^q_{ij})_{L}(\delta^q_{j})_{LR}(\delta^q_{ji})_{R}
  \right],
\end{eqnarray}
where $\alpha_s = g_s^2/(4\pi)$, and
$m_{\tilde{g}}$ and $\overline{m}_{\tilde{q}}$ 
are the gluino and averaged squark masses.
In Table \ref{tab:CEDM}, we show the  
constraints on the mass insertion parameters from the CEDM bounds.
Here we use the formulae in Ref. \cite{Hisano:2004pw},
and SUSY parameters are set to be $m_{\rm SUSY} = 500 \ \mbox{GeV}$
and $\tan \beta = 10$.

\begin{table}[t]
  \begin{center}
    \caption{{\footnotesize 
        Constraints on the mass insertion parameters of squarks
        from the EDMs of the neutron and the $^{199}$Hg atom. 
        Here, we evaluate the gluino diagram
        contribution to them and require them to be smaller than the
        experimental bounds.
        The superparticle masses are set to be $m_{\rm SUSY}$.  
        The bounds on the
        combination of the mass insertion parameters in this table are
        proportional to $m_{\rm SUSY}^2$. 
        Here we consider only diagrams
        proportional to $\tan\beta$ for the CEDMs of the strange and down
        quarks.}}
    \label{tab:CEDM}
    \vspace{0.5cm}
    \begin{tabular}{|c|c||c|c|}
      \hline
      \multicolumn{4}{|c|}
      {$^{199}$Hg EDM (neutron EDM) 
        ($m_{\rm SUSY}=500$ GeV and $\tan\beta=10$)}
      \\ \hline
      Im$[(\delta^u_{12})_{R}(\delta^u_{21})_{L}]$                       
      &  
      $0.8(1)\times 10^{-3}$ &
      Im$[(\delta^u_{13})_{R}(\delta^u_{31})_{L}]$ 
      &  
      $3(5)\times 10^{-6}$ 
      \\ \hline
      Im$[(\delta^d_{12})_{R}(\delta^d_{21})_{L}]$                       
      &  
      $0.6(1)\times 10^{-3}$ &
      Im$[(\delta^d_{13})_{R}(\delta^d_{31})_{L}$] 
      &  
      $2(4)\times 10^{-5}$ 
      \\ \hline
      Im$[(\delta^d_{23})_{R}(\delta^d_{32})_{L}$]                       
      &  
      $3(0.2)\times 10^{-3}$ &
      &  
      \\ \hline
    \end{tabular}
  \end{center}
\end{table}

The bound on $(\delta^u_{13})_{L,R}$ in the up-type squark mass
matrices comes from the up quark CEDM, and it severely constrains SUSY
models because of the enhancement by the large top quark mass. 
However, the up quark CEDM is suppressed 
below the experimental bound
when the mediation of the SUSY breaking to our visible sector
is flavor-blind. 
The flavor
violating mass terms for the squarks are radiatively generated
by the Yukawa couplings. 
On the other hand, 
if the flavor universality is violated at tree level as in the
orbifold GUT models, the up quark CEDM gives a stringent
constraint on the models, as will be shown.

Let us briefly review the realization of realistic flavor structures of 
fermions and sfermions in the context of orbifold GUT models
where both the GUT symmetry and SUSY are
broken by the boundary conditions in extra dimensions. 
Especially we consider one construction of a SUSY SU(5) GUT in 
five dimensions proposed by Hall and Nomura \cite{Hall:2002ci}.
The five-dimensional spacetime is postulated to be factorized 
into the usual four-dimensional
Minkowski spacetime and the extra spatial dimension 
compactified on an $S^1/Z_2$ orbifold.
The compactification is realized by identifying the fifth coordinate
$y$ under the reflection ${\cal Z}:\ y \to - y$ and 
the translation ${\cal T}:\ y \to y + 2 \pi R$.
The physical space is a line interval, $0 \leq y \leq \pi R$,
which has two branes at the orbifold fixed points $y=0$ and $y= \pi R$.
The reduction of five-dimensional SUSY into 
four-dimensional ${\cal N}=1$ SUSY is attributed to
the $y$-reflection ${\cal Z}$,
and the SU(5) breaking comes from 
the translation ${\cal T}$ with the non-trivial action of 
$P={\rm diag} (+,+,+,-,-)$ on a fundamental multiplet.
We will see that the appropriate choice of the boundary conditions
leads to the ${\cal N}=1$ SUSY SM.

The five-dimensional SU(5) vector multiplet lives in the bulk,
and consists of a five-dimensional vector boson $A_M$,
gauginos $\lambda$ and $\lambda'$, and a real scalar $\sigma$.
In a four-dimensional ${\cal N}=1$ SUSY viewpoint,
the vector multiplet is rewritten by a four-dimensional vector
superfield $V=(A_\mu, \lambda)$ and an adjoint chiral superfield 
$\Sigma=((\sigma+iA_5)/\sqrt{2},\lambda')$.
The boundary conditions on the gauge multiplet are
\begin{eqnarray}
  \left( 
    \begin{array}{c}
      V^{(p)} \\ \Sigma^{(p)}
    \end{array}
  \right)(x^\mu, y) = 
  \left( 
    \begin{array}{c}
      V^{(p)} \\ - \Sigma^{(p)}
    \end{array}
  \right)(x^\mu, - y) = 
  p \left( 
    \begin{array}{c}
      V^{(p)} \\ \Sigma^{(p)}
    \end{array}
  \right)(x^\mu, y+2 \pi R),
\end{eqnarray}
where $p$ stands for eigenvalues of the matrix $P$.
Since $p = 1$ for the SU(3), SU(2), and U(1) components
and $p= -1$ for the others,
the SU(5) symmetry is preserved in the bulk and on the brane at $y=0$,
while only the SM gauge symmetry survives on the brane at $y=\pi R$.
In the four-dimensional viewpoint,
there appear 
infinite Kaluza-Klein (KK) towers of 
the partner states with identical quantum charges.
Due to our choice of the boundary conditions,
only the vector superfields with $p=1$ have zero modes
which are correctly identified with the vector multiplets
in the SUSY SM,
whereas the vector 
superfields with $p=-1$ and the adjoint chiral superfield 
have no zero mode.

One five-dimensional hypermultiplet, which contains
two complex scalars $\phi$ and $\phi^c$, and two chiral fermions $\psi$
and $\psi^c$,
is written in terms of two ${\cal N}=1$ four-dimensional 
chiral multiplets, $\Phi=(\phi, \psi)$ and $\Phi^c=(\phi^c,\psi^c)$.
The boundary conditions are
\begin{eqnarray}
  \left( 
    \begin{array}{c}
      \Phi^{(p)} \\ \Phi^{c(p)}
    \end{array}
  \right)(x^\mu, y) = 
  \left( 
    \begin{array}{c}
      \Phi^{(p)} \\ - \Phi^{c(p)}
    \end{array}
  \right)(x^\mu, - y) = 
  p \ \eta_\Phi \left( 
    \begin{array}{c}
      \Phi^{(p)} \\ \Phi^{c(p)}
    \end{array}
  \right)(x^\mu, y+2 \pi R),
\end{eqnarray}
where $\eta_\Phi = 1$ or $-1$.
In the four-dimensional viewpoint,
only the components with $p\ \eta_\Phi = 1$ 
in $\Phi$ contain their zero modes
while the other modes in $\Phi$ and $\Phi^c$ have masses of order of
the compactification scale, as in the case of the vector multiplet.

The main purpose of the orbifold GUTs is to solve
the notorious doublet-triplet splitting problem
by invoking the boundary conditions.
When the fundamental and anti-fundamental Higgs hypermultiplets, 
$(H,H^c)$ and $(\bar{H},\bar{H}^c)$, have properties 
of $\eta_H = \eta_{\bar{H}}=-1$,
only the doublet components of 
the KK towers of $H$ and $\bar{H}$ contain zero modes.
With the help of U(1)$_R$ symmetry,
the the zero mode Higgs doublets are obliged to be massless.
The U(1)$_R$ symmetry also eliminates
dimension four and five proton decays.

A successful gauge coupling unification is achieved
at the cutoff scale $M$,
although the SU(5) symmetry is explicitly broken on the SM brane.
Let us verify that the large volume factor dilutes 
the SU(5) breaking effects,
in the viewpoint of the four-dimensional effective theory.
The SM gauge coupling constants are consists of 
the five-dimensional SU(5)-bulk gauge coupling constant $g_5$ and 
those from both brane-localized kinetic terms $\tilde{g}_a$,
and written 
\begin{eqnarray}
  \frac{1}{g_a^2} = \frac{\pi R}{g_5^2} + \frac{1}{\tilde{g}_a^2}.
\end{eqnarray}
Here $a=1,2,3$ labels the U(1)$_Y$, SU(2)$_L$ and SU(3)$_C$ gauge groups,
respectively.
The existence of the SU(5)-breaking brane would prevent us from making
any prediction
without the knowledge of ultraviolet physics.
This problem is easily avoided by the strong coupling hypothesis.
Suppose that the gauge interaction is as strong at the cutoff scale
as perturbative expansions break down.
Requiring that all loop-expanded diagrams are equally weighted,
we obtain $g_5^2 \simeq 16 \pi^3 /C M$ and $g_5^2 \simeq 16 \pi^2 /C_a$, 
where $C's$ represent group theoretical factors.
Given large volume of the extra dimension such as $MR \sim O(100)$,
we can ignore the contribusions from the SU(5)-breaking brane 
and the bulk gauge coupling constant should be
$1/g_*^2 \equiv \pi R/g_5^2 \simeq 2$ for obtaining $g_a \simeq 0.7$.
The largeness of the extra dimension leaves room
for sizable deviation from the conventional MSSM
prediction of the QCD coupling,
in which threshold corrections from KK particles 
above the compactification scale enters in an SU(5)-violating manner.
Since the leading power-law contributions to the gauge couplings are 
SU(5)-symmetric,
the renormalization group (RG)
flows of the differences among the gauge coupling constants
are logarithmic and under control.
The parameter set
as $R^{-1} \sim 10^{15}$ GeV and $M \sim 10^{17}$ GeV,
for example,
reconciles the 
the prediction of the orbifold GUT with the experimental value of 
the QCD coupling constant \cite{Hall:2002ci}.

Supersymmetry breaking leading to soft breaking terms in
the MSSM is also attributed 
to the boundary conditions.
One of the most attractive features of 
such a boundary condition breaking is 
that a single parameter $\alpha$ describes all the soft terms.
The supersymmetry breaking arises from the following twisted
boundary conditions for SU(2)$_R$ doublets:
\begin{eqnarray}
  \left( 
    \begin{array}{c}
      \lambda^{(p)} \\ \lambda^{\prime(p)}
    \end{array}
  \right)(x^\mu, y) = 
  \left( 
    \begin{array}{c}
      \lambda^{(p)} \\ - \lambda^{\prime(p)}
    \end{array}
  \right)(x^\mu, - y) = 
  e^{2 \pi i \alpha \sigma_2} p \left( 
    \begin{array}{c}
      \lambda^{(p)} \\ \lambda^{\prime(p)}
    \end{array}
  \right)(x^\mu, y+2 \pi R),
\end{eqnarray}
and
\begin{eqnarray}
  \left( 
    \begin{array}{c}
      \phi^{(p)} \\ \phi^{c(p)\dag}
    \end{array}
  \right)(x^\mu, y) = 
  \left( 
    \begin{array}{c}
      \phi^{(p)} \\ - \phi^{c(p)\dag}
    \end{array}
  \right)(x^\mu, - y) = 
  e^{2 \pi i \alpha \sigma_2} p \ \eta_\Phi \left( 
    \begin{array}{c}
      \phi^{(p)} \\ \phi^{c(p)\dag}
    \end{array}
  \right)(x^\mu, y+2 \pi R).
\end{eqnarray}
Given an appropriate choice of the value of
$\widetilde{m} \equiv \alpha R^{-1}$,
we can reproduce the weak-scale SUSY scenarios:
$\alpha \sim 10^{-13}$ for $R^{-1} \sim 10^{15}$ GeV for instance.
The MSSM gaugino masses, which arise at the compactification scale,
are given by
\begin{eqnarray}
  M_a = \frac{g_a^2}{g_*^2}\widetilde{m}.
\end{eqnarray}
Notice that the usual gaugino mass unification is
retained as in the case of the conventional SUSY GUTs.
The resulting soft SUSY breaking terms 
for scalar zero modes in the orbifold GUT model is
\begin{eqnarray}
  - {\cal L}_{\rm soft} & = & \widetilde{m}^2 
  (|h|^2 + |\phi_B|^2) 
  - \widetilde{m} [3 f \phi_B \phi_B h + 2 f \phi_B \phi_b h
  + f \phi_b \phi_b h + {\rm h.c.}],
\end{eqnarray}
where $h$ and $\phi_{B(b)}$ denote the Higgs doublets and
bulk (brane) matter fields collectively,
and $f$ the corresponding Yukawa coupling constants 
in the superpotential.
We notice that the SUSY breaking is related with bulk fields
and dictated by $\widetilde{m}$.

\begin{figure}[t]
  \begin{center} 
    \scalebox{.4}{\includegraphics*{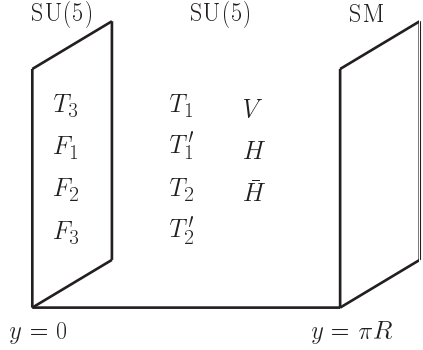}}
    \caption{{\footnotesize
          Geometric structure and field configuration in the 
          fifth dimension $y$.
          The SU(5) vector multiplet $V$, 
          the fundamental (anti-fundamental) Higgs hypermultiplet 
          $H(\bar{H})$ and four ten-plet matter hypermultiplets 
          $T_{1,2},T_{1,2}^\prime$ live in the bulk.
          One ten-plet matter chiral superfield $T_3$ and
          three five-plet matter chiral superfields $F_{1,2,3}$
          are located on the SU(5) symmetric brane at $y=0$.
          There are no localized fields on the SM brane at $y=\pi R$.
    }}
  \label{fig:orbifold}
  \end{center}
\end{figure}

Let us consider the matter configuration 
and its consequences for the flavor structures of fermions and sfermions.
There are three options for locating a matter field:
in the bulk, on the SU(5) symmetric brane at $y=0$ 
or on the SM brane at $y=0$.
Here we employ the following prescription suggested by Hall and Nomura:
the first two families of ten-plets $T_{1,2},T_{1,2}^\prime$ with
$\eta_{T_{1,2}} = - \eta_{T_{1,2}^\prime} = 1$
live in the bulk,
while the third generation ten-plet $T_3$ and 
all the three families of the five-plet matters $F_{1,2,3}$
are located on the SU(5) symmetric brane.
The first two generations of the up-type quark singlets
$\bar{U}_{1,2}$ and the charged-lepton singlets $\bar{E}_{1,2}$
in the MSSM 
are identified with the zero modes of $T_{1,2}$,
while those of the quark doublets $Q_{1,2}^{\prime}$ with those of
$T_{1,2}^\prime$.
In Figure \ref{fig:orbifold} we schematically depict
the geometric structure and the field configuration.
The locations of the matter multiplets are determined by the combined 
results given by the phenomenological analysis below.

First note that the matter configuration in Figure \ref{fig:orbifold}
accounts for the hierarchical quark and lepton masses and mixings
by a virtue of the volume suppression.
The Yukawa interactions among the matter and Higgs multiplets
are allowed only on the SU(5) brane.
The resulting Yukawa couplings in the MSSM are given by
\begin{eqnarray}
  W_Y & \sim & (Q^\prime_1, Q^\prime_2, Q_3)
  \left(
    \begin{array}{cc|c}
      \epsilon^2 & \epsilon^2 & \epsilon \\ 
      \epsilon^2 & \epsilon^2 & \epsilon \\ \hline
      \epsilon & \epsilon & 1 
    \end{array}
  \right)
  \left(
    \begin{array}{c}
      \bar{U}_1 \\ \bar{U}_2 \\ \bar{U}_3
    \end{array}
  \right)
  \nonumber \\
  && + (Q^\prime_1, Q^\prime_2, Q_3)
  \left(
    \begin{array}{ccc}
      \epsilon & \epsilon & \epsilon \\ 
      \epsilon & \epsilon & \epsilon \\ \hline
      1 & 1 & 1 
    \end{array}
  \right)
  \left(
    \begin{array}{c}
      \bar{D}_1 \\ \bar{D}_2 \\ \bar{D}_3
    \end{array}
  \right)
  + (\bar{E}_1, \bar{E}_2, \bar{E}_3)
  \left(
    \begin{array}{ccc}
      \epsilon & \epsilon & \epsilon \\ 
      \epsilon & \epsilon & \epsilon \\ \hline
      1 & 1 & 1 
    \end{array}
  \right)
  \left(
    \begin{array}{c}
      L_1 \\ L_2 \\ L_3
    \end{array}
  \right)
  \label{eq:Yukawa}
\end{eqnarray}
with coefficients of order of unity.
The large extra dimension whose size is of order of $R \sim 100 M^{-1}$
suffices for the mass hierarchy,
so that $\epsilon \simeq (MR)^{-1/2} \sim 0.1$.
Only the $(3,3)$ element in the up-type Yukawa couplings 
and the third rows of the down-type and charged-lepton Yukawa couplings
preserves SU(5) in the conventional sense,
because the zero modes of the first two generations of the ten-plets
stem from different bulk multiplets.
Therefore the up-type quark Yukawa matrix is not symmetric and
the masses of the first two generations of the down-type quarks 
are not associated with the counterparts of charged-leptons
while retaining the bottom and tau mass unification.
We should admit that the simple geometric structure of the orbifold GUT 
does not explain the hierarchy within the first two generations.
There are several ways to avoid this difficulty.
One of the options is to impose an additional U(1) flavor symmetry 
with the charge assignments such as
$T_1^{(\prime)}(1), T_{2,3}^{(\prime)}(0), F_{1,2,3}(1)$.
Given the U(1) flavor symmetry breaking parameter of order of $\epsilon$,
more realistic Yukawa coupling matrices are obtained as
\begin{eqnarray}
  f_u  & \sim & \left(
    \begin{array}{cc|c}
      \epsilon^4 & \epsilon^3 & \epsilon^2 \\ 
      \epsilon^3 & \epsilon^2 & \epsilon \\ \hline
      \epsilon^2 & \epsilon & 1 
    \end{array}
    \right), \quad
  f_d \sim \left(
    \begin{array}{ccc}
      \epsilon^3 & \epsilon^3 & \epsilon^3 \\ 
      \epsilon^2 & \epsilon^2 & \epsilon^2 \\ \hline
      \epsilon & \epsilon & \epsilon
    \end{array}
    \right), \quad
  f_e \sim \left(
    \begin{array}{ccc}
      \epsilon^3 & \epsilon^3 & \epsilon^3 \\ 
      \epsilon^2 & \epsilon^2 & \epsilon^2 \\ \hline
      \epsilon & \epsilon & \epsilon
    \end{array}
    \right),
  \label{eq:realYukawa}
\end{eqnarray}
instead of Eq. (\ref{eq:Yukawa}).
The mass hierarchy between the top and bottom quarks 
is also explained for a small $\tan \beta$ region.
In the rest of this letter,
we postulate that some mechanism works to reproduce
the realistic type of the Yukawa structure of Eq. (\ref{eq:realYukawa}).
By introducing three generations of right-handed neutrino fields, 
the smallness of the neutrino masses are explained via 
the seesaw mechanism \cite{seesaw}. 
We can place the right-handed neutrinos either in the bulk
or in the SU(5) symmetric brane.
In both cases, the Dirac Yukawa couplings and the Majorana mass terms
for the right-handed neutrinos are given in the SU(5) brane.
Since three generations of the left-handed leptons $L_{1,2,3}$
have same geometric profiles,
the large mixing angles in the neutrino sector are naturally realized.

The gauge coupling unification in the orbifold GUTs requires 
a lower compactification
scale than the usual GUT scale in the conventional SUSY GUT scenario.
Thus we might 
expect that dimension six nucleon decays mediated by the 
heavy $X$ gauge bosons with mass of order of $R^{-1}$ are dangerous.
Here we recall that $\bar{U}_{1,2},\bar{E}_{1,2}$ and
$Q_{1,2}^{\prime}$ in the MSSM originate from different 
five-dimensional ten-plets
and that the $X$ gauge bosons cannot convert $Q_{1,2}^{\prime}$ into 
$\bar{U}_{1,2}$ nor $\bar{E}_{1,2}$.
Therefore the proton decay rates are suppressed to 
an acceptable level by small flavor mixings 
between the first two and third generations.

The same matter configuration considerably ameliorates
the SUSY flavor problem.
The boundary condition SUSY breaking 
ensures universal soft scalar
masses for bulk fields 
with vanishing soft scalar masses for brane fields, 
as we mentioned above.
The geometry provides the following soft terms for the zero modes:
\begin{eqnarray}
  - {\cal L}_{\rm soft}
  & = & \widetilde{m}^2 [\tilde{q}^* P_T \tilde{q}
  + \tilde{u^c}^* P_T \tilde{u^c}
  + \tilde{e^c}^* P_T \tilde{e^c}] \nonumber \\
  && 
  - \widetilde{m} [\tilde{q} (f_u + P_T f_u + f_u P_T) \tilde{u}^c h_u 
  + \tilde{q} (f_d + P_T f_d) \tilde{d}^c h_d  \nonumber \\
  && \qquad + \tilde{e^c} (f_e + P_T f_d) \tilde{l} h_d
  + {\rm h.c.}], \nonumber \\
  P_T & = & \left(
  \begin{array}{ccc}
     1 & 0 & 0 \\
     0 & 1 & 0 \\
     0 & 0 & 0
  \end{array}
  \right),
\end{eqnarray}
at the compactification scale.
At the electroweak scale we obtain universal masses 
for left-handed sleptons and right-handed down-type squarks,
while the third generations of left-handed quark doublets,
right-handed up-type quarks and charged sleptons
are different from their first two generations in mass.
The mass degeneracy especially within the first two generations
adequately suppresses
the most dangerous flavor violating processes
caused by superparticles,
such as the $K$ meson mixing and the $\mu \to e \gamma$ decay.
Sizable flavor violating effects are originated only 
from the separation of the locations of the ten-plets.
In other words it is worth investigating implications
from flavor mixings between 
the first two generations and the third generation in 
left-handed squark,
right-handed up-type squark and right-handed slepton sectors.

Let us explore experimental signatures from the remarkable 
flavor structure realized in the orbifold GUT models.
The different location of the third generation ten-plet $T_3$ 
in the extra dimension 
from those of the first two generations induces 
the sfermion flavor mixing for left-handed squarks, 
right-handed up-type squarks and charged sleptons.
Such a situation is reminiscent of the conventional SUSY SU(5) GUTs.
In the SUSY SU(5) GUTs with the minimal supergravity boundary condition
the renormalization group (RG) effects above the GUT scale 
generate similar flavor-violating terms for the sfermions
via the top and bottom Yukawa couplings \cite{SUSY_GUT_flavor}. 
We focus on the CEDM of the up quark originated from the up-type squark 
mixings and LFV from the right-handed slepton sector.
We will show that the observations of hadronic EDMs
and the LFV $\mu \to e \gamma$ decay have a great 
potential to distinguish the two frameworks.

We can estimate the magnitude of the up quark CEDM using the mass
insertion approximation.
The Yukawa matrices are related to those in the mass eigenbasis 
by the following biunitary transformation:
$f_u = V_{u_L}^T \hat{f}_u V_{u_R}$,
$f_d = V_{d_L}^T \hat{f}_d V_{d_R}$,
$f_e = V_{e_R}^T \hat{f}_e V_{e_L}$.
Then the soft terms in the super-CKM basis are 
\begin{eqnarray}
  - {\cal L}_{\rm soft}
  & = & \widetilde{m}^2 [\tilde{q}^* C_{u_L} \tilde{q}
  + \tilde{u^c}^* C_{u_R} \tilde{u^c}
  + \tilde{e^c}^* C_{e_R} \tilde{e^c} \nonumber \\
  && 
  - \widetilde{m} 
  [\tilde{q} (\hat{f}_u + C_{u_L}^* \hat{f}_u + \hat{f}_u C_{u_R}) 
  \tilde{u}^c h_u 
  + \tilde{q} (V^* \hat{f}_d 
  + C_{u_L}^* V^* \hat{f}_d) \tilde{d}^c h_d \nonumber \\
  && \quad + \tilde{e^c} (\hat{f}_e + C_{e_R}^* \hat{f}_e) \tilde{l} h_d
  + {\rm h.c.}],
\end{eqnarray}
where $V = V_{u_L} V_{d_L}^*$ is the KM matrix.
Additional flavor violating effects are controlled by 
\begin{eqnarray}
  (C_f)_{ij} = (V_f P_T V_f^\dag)_{ij} 
  = {\bf 1}_{ij} - (V_f)_{i3} (V_f^*)_{j3}, \quad f = u_L, u_R, e_R.
\end{eqnarray}
In generic CP violating off-diagonal elements
in the left-handed up-type squark mass matrix
are not related to the right-handed counterparts.
Therefore we expect too large CP violation as
$\mbox{Im}[(\delta_{13}^u)_L (\delta_{31}^u)_R] 
\sim \epsilon^4 \sim 10^{-4}$.
Although the squark mixings are diluted by the gluino focusing effect,
the CEDM of the up quark is still dangerous 
like the $\mu \to e\gamma$ decay.

We have performed numerical calculations clarifying the observable
hadronic EDMs are indeed induced by the CP-violating sfermion mixings
as well as LFV processes.
Since there is a sizable mass splitting among the sfermions,
we work in the basis where the sfermion mass matrices are fully 
diagonalized,
instead of the mass insertion approximation.
The CEDM of the up quark is given by
\begin{eqnarray}
  d^C_u = c \frac{\alpha_s}{4 \pi} \frac{1}{M_3} \sum_{I=1}^6
  \left\{ \left( - \frac{1}{3} G_1 (x_I) - 3 G_2 (x_I) \right)
    \mbox{Im} \left[ (U_u^\dag)_{1I} (U_u)_{I4} \right]
    \right\},
\end{eqnarray}
where $(U_u)_{Ii}$ relates
the up-type squarks $\tilde{u}^{(m)}_I$ with mass $m_I$ to
those in a flavor basis $(\tilde{u}_L^{(f)}, \tilde{u}_R^{(f)})_i$
by the following unitary transformation: 
$\tilde{u}^{(m)}_I = (U_u)_{Ii} (\tilde{u}_L^{(f)},\tilde{u}_R^{(f)})_i$.
Here $c\sim 0.9$ is the QCD correction factor at low energy, 
$x_I (\equiv m_I^2/M_3^2)$ parametrizes mass ratios of the up-type
squarks to the gluino,
and the loop functions $G_1(x)$ and $G_2(x)$ are defined by
\begin{eqnarray}
  G_1(x) \equiv \frac{1 - x^2 + 2x  \ln x}{2(x - 1)^3}, \quad
  G_2(x) \equiv \frac{1 - 4x + 3x^2 - 2x^2  \ln x}{2(x - 1)^3},
\end{eqnarray}
where $G_1(1) = - 1/6, G_2(1) = - 1/3$.

\begin{figure}[t]
  \begin{center}
    \hspace*{-0.5cm}
    \scalebox{.65}{\includegraphics*{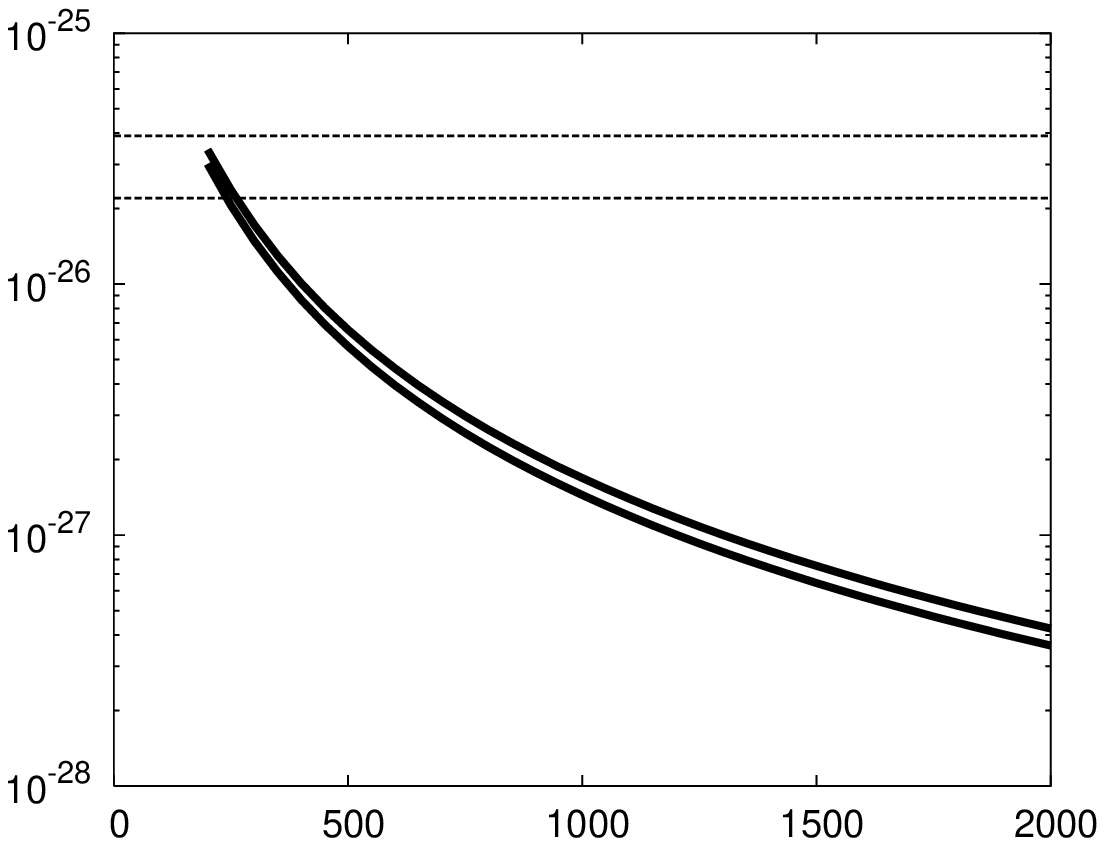}}
    \put(-160,170){{\bf CEDM of up quark}}
    \put(-230,70){\rotatebox{90}{$d^C_u\ [\mbox{cm}]$}}
    \put(-115,-15){$\widetilde{m}$ [GeV]}
    \put(-110,-35){(a)}
    \put(-90,70){{\footnotesize $\tan \beta = 10$}}
    \put(-120,50){{\footnotesize $\tan \beta = 3$}}
    \put(-100,140){{\footnotesize Neutron EDM}}
    \put(-100,115){{\footnotesize Hg EDM}}
    \scalebox{.65}{\includegraphics*{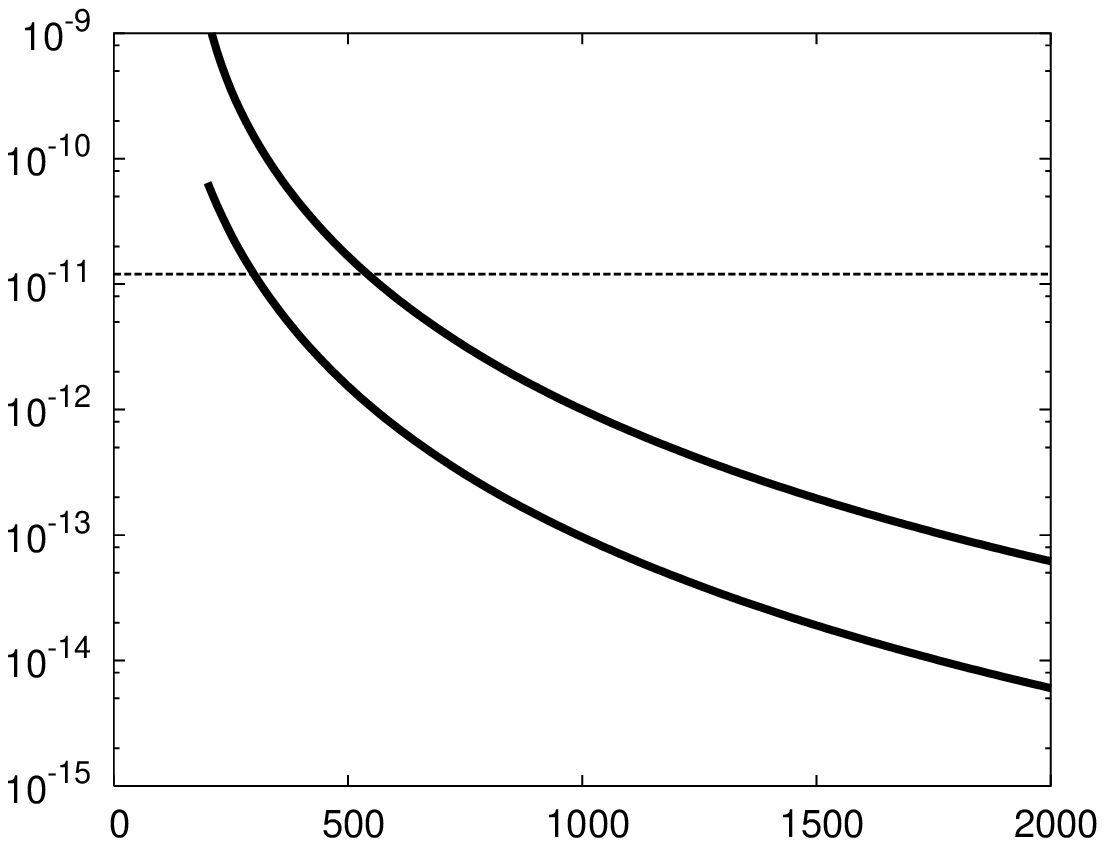}}
    \put(-180,170){{\bf Branching ratio for $\mu \to e \gamma$}}
    \put(-230,60){\rotatebox{90}{Br$(\mu \to e \gamma)$}}
    \put(-115,-15){$\widetilde{m}$ [GeV]}
    \put(-105,-35){(b)}
    \put(-70,80){{\footnotesize $\tan \beta = 10$}}
    \put(-150,50){{\footnotesize $\tan \beta = 3$}}
    \caption{{\footnotesize 
        CEDM of up quark in (a) 
        and branching ratio for $\mu \to e \gamma$ decay in (b)
        as functions of the SUSY breaking parameter $\widetilde{m}$.
        We set the magnitudes of the flavor violating parameters to
        be $|(C_f)_{13}| = 0.01, |(C_f)_{23}| = 0.04$,
        and the five-dimensional gauge coupling constant to be
        $1/g^2_* = 1.9$ at 
        $R^{-1} = 1 \times 10^{14}\ \mbox{GeV}$, 
        and postulate maximal CP violation.
        For SUSY parameters we take 
        $\tan \beta = 3, 10$ and $\mu < 0$.
        The present experimental upper bounds are
        $d_u^C < 2.2 \times 10^{-26}$ cm (Hg EDM),
        $d_u^C < 3.9 \times 10^{-26}$ cm (Neutron EDM),
        and Br$(\mu \to e \gamma) < 1.2 \times 10^{-11}$.
      }}
    \label{fig:prediction}
  \end{center}
\end{figure}

In Figure \ref{fig:prediction}
we show the predicted values of the CEDM of the up quark in (a)
and the branching ratio for the LFV $\mu \to e \gamma$ decay in (b)
as functions of $\widetilde{m}$ for 
$\tan \beta = 3,10$.
Here we set the magnitudes of 
sfermion mixing parameters to $|(C_f)_{13}|=0.01$ and $|(C_f)_{23}|=0.04$
at the compactification scale $R^{-1} = 1 \times 10^{14}$ GeV, and 
postulate maximal CP violation.
These values are naturally inferred from the corresponding
CKM matrix elements, because the rotation of the ten-plets 
in the generation space simultaneously 
dictates the hierarchical structure of the CKM matrix and the additional
sfermion mixings.
The five-dimensional gauge coupling constant is taken to be 
$1/g^2_* = 1.9$, and the RG effects from the compactification scale
to the $Z$-boson mass scale
are included.
The Higgs mixing parameter $\mu$ is chosen to be negative,
which is preferred by considering the precise bottom-tau unification 
\cite{Hall:2002ci}.
The dependence of the signature of $\mu$ on 
the predicted values is weak.
The present experimental bounds are 
$d_u^C < 2.2 \times 10^{-26}$ cm (Hg EDM),
$d_u^C < 3.9 \times 10^{-26}$ cm (Neutron EDM),
and Br$(\mu \to e \gamma) < 1.2 \times 10^{-11}$ \cite{Brooks:1999pu}.
Due to the gluino focusing effect
the mass insertion parameters in the up-type squarks 
at the superparticle mass scale are diminished as
\begin{eqnarray}
  (\delta^u_{ij})_{L/R}|_{m_{\rm SUSY}} \sim 
  \left. \frac{(\delta^u_{ij})_{L/R}}{1 + \eta (g_3^4/g_*^4)} 
  \right|_{R^{-1}}, 
    \quad \eta \sim 5,
\end{eqnarray}
and their dependence on $R$ is only logarithmic,
while those in the slepton sector are insensitive to RG effects.
Notice that 
in our setup various observables,
such as flavor violating processes and mass spectra,
are uniquely determined 
by a single parameter $\widetilde{m}$,
once the flavor mixing parameters are fixed.
These figures demonstrate that
the orbifold GUT scenario marginally survives
the bounds from the hadronic EDMs and the $\mu \to e \gamma$ decay.

A new deuteron EDM experiment is proposed aiming for the sensitivity 
of $d_D \sim 10^{-27}\ e$ cm \cite{Semertzidis:2003iq},
which is converted to the CEDM bounds as
$d_u^C \sim 10^{-28}$ cm,  
$d_d^C \sim 10^{-28}$ cm and  
$d_s^C \sim 10^{-26}$ cm.
Furthermore the MEG experiment will reach to
the level of Br$(\mu \to e \gamma) \sim 10^{-14}$ in the near future
\cite{MEG}.
These experiments will cover the interesting parameter region 
in the orbifold GUT models.
Null signature of the deuteron EDM or the $\mu \to e \gamma$ decay would
oblige us to change the matter configuration in the extra dimension.

Let us discuss other possible choices of matter configurations
in the framework of the orbifold GUTs, 
where SUSY is broken by the boundary conditions.
We can place $T_3$ and $F_{1,2,3}$ on the SM brane
instead of the SU(5) symmetric one.
The gross flavor structures of fermions and sfermions resemble those
in the Hall--Nomura model, 
and thus similar flavor violating effects are derived.
Future improvements of the experiments
may force us to eliminate the off-diagonal elements 
in the sfermion mass matrices
if the signatures are not observed.
This means that all of the ten-plets should be located on the same place
in the light of the quark mixings.
The first two generations of ten-plets $T_{1,2}$ are forbidden 
from living in the SU(5) brane 
in order to suppress the dimension six proton decays, 
while $T_3$ should be localized on one of the 
two branes to reproduce the large top quark mass.
Thus $T_{1,2,3}$ must be located on the SM brane
\cite{Hebecker:2001wq}.
The locations of the three generations of five-plets are 
either in the bulk or on the SM brane 
in order to allow the Yukawa interactions.
In the bulk five-plet case, 
the ratio of the bottom to top Yukawa couplings
is suppressed by the large volume of the extra dimension, 
and the soft scalar masses of the five-plets are lifted by 
$\widetilde{m}$.
In the case where the five-plets are located on the SM brane,
we obtain the same Yukawa structures as those in the MSSM.
In both cases, 
the SU(5) relation for the Yukawa couplings is discarded,
and we need some other mechanism to realize
the flavor hierarchy in the fermion masses and mixings.
As far as non-universal configuration for the matter multiplets is
responsible for the sfermion mixings, 
which are dictated by the corresponding CKM matrix elements,
not only LFV processes but also the hadronic EDMs is predicted to
be just below the current limits.
This conclusion is generic, and not altered even in extended 
models possessing higher-dimensional spacetimes or 
larger gauge groups.

The predictions of the orbifold GUTs should be compared to
those of four-dimensional SUSY SU(5) GUTs.
If the sfermion masses arise in a flavor-blind manner 
from the dynamics above the GUT scale
(for example, the minimal supergravity scenario),
the sfermion flavor mixings are induced only by
the RG running effects from the cutoff scale $M_G$ 
and thus they are loop-suppressed.
In SUSY SU(5) GUTs 
there exist the color triplet partners of the MSSM Higgs doublets. 
The Yukawa interactions mediated by the Higgs triplets 
with mass $M_{H^c}$
lead to flavor mixings in the ten-plet scalars.
In the minimal SU(5) model without heavy right-handed neutrinos,
the resulting flavor mixings are controlled by the CKM matrix
and the top and bottom Yukawa couplings $f_{t,b}$ as
\begin{eqnarray}
(m_{{\tilde{u}_L}}^2)_{ij}  &\simeq&-
V_{i3}V_{j3}^*  \frac{f_{b}^2}{(4\pi)^2}
(3m_0^2+ A_0^2)
\left( 2 \ln \frac{M_G^2}{M_{H_c}^2}+ 
\ln \frac{M_{H_c}^2}{m^2_{\rm SUSY}} \right),
\nonumber\\
(m_{\tilde{u}_R}^2)_{ij}  &\simeq& -
e^{-i\varphi_{ij}} 
V_{i3}^* V_{j3} \frac{2f_{b}^2}{(4\pi)^2}
(3m_0^2+ A_0^2)
\ \ln \frac{M_G^2}{M_{H_c}^2}, \nonumber\\
(m_{{\tilde{d}_L}}^2)_{ij}  &\simeq&-
V_{3i}^*
V_{3j} \frac{f_{t}^2}{(4\pi)^2} 
(3m_0^2+ A_0^2)
\left( 3 \ln \frac{M_G^2}{M_{H_c}^2}
+ \ln\frac{M_{H_c}^2}{m^2_{\rm SUSY}} \right),
\nonumber\\
(m_{\tilde{d}_R}^2)_{ij}  &\simeq& 0, \nonumber\\
(m_{\tilde{e}_L}^2)_{ij}  &\simeq& 0, \nonumber\\
(m_{{\tilde{e}_R}}^2)_{ij}  &\simeq&-
V_{3i}V^*_{3j}
\frac{3 f_{t}^2}{(4\pi)^2} 
(3m_0^2+ A_0^2) \
\ln \frac{M_G^2}{M_{H_c}^2},
\end{eqnarray}
with $i\ne j$, where
$\varphi_{ij} \equiv \varphi_i - \varphi_j$
parametrizes the GUT-origin phases independent of the KM phase.
Here $m_0$ and $A_0$ are the universal soft 
scalar mass and the trilinear coupling, respectively. 
Although both the left-handed and right-handed up-type squarks have the
off-diagonal components with different CP phases,
the induced CEDM of the up quark is suppressed by the fourth power of
bottom Yukawa coupling in addition to the loop suppression
\cite{Romanino:1996cn}
\footnote{
The radiative corrections to the off-diagonal terms for left- and
right-handed up squarks may be generated by the top quark Yukawa
coupling, when the misalignment between the Yukawa couplings
in the MSSM and the GUTs is introduced, such as by the higher
dimensional operators \cite{higher_dim_op}. 
However, the up quark CEDM is still
expected to be smaller than $\sim 10^{-28}$ cm since the correction is
suppressed by loop factors.
}.
For example, 
we obtain
$d^C_u \sim 10^{-28}\ \mbox{cm}$ for $m_{\rm SUSY} = 500$ GeV and
$\tan \beta \simeq 35$.
Even for a large $\tan \beta$ region
the predicted hadronic EDMs are well below the current upper limits.
If we observe the deuteron EDM above the $10^{-27}\ e$ cm level,
it would be attributed to the down or strange quark 
contribution coming from other physics in the context of 
four-dimensional GUTs.
On the other hand, the right-handed slepton mixings involve the large
top Yukawa coupling, so that LFV decay processes are 
within the scope of future experiments,
although the various contributions tend to cancel out
\cite{Hisano:1996qq}.
We emphasize that by combining the results of the hadronic EDM
and LFV searches 
we can distinguish the orbifold framework from the conventional SU(5) 
SUSY GUTs.

It is also important to measure various hadronic EDM processes,
because the light quark CEDM contributions to hadronic EDMs are 
quite different.
By investigating the correlation of the observed hadronic EDMs,
we might find which contribution dominates over.
If we identify the origin,
it would give us crucial hints for the physics beyond the SM.

In this letter we have investigated the hadronic EDM constraints
on the orbifold GUT framework, which explains the 
fermion and sfermion flavor structures
due to an appropriate choice of the locations of the matter fields in 
extra dimensions.
For definiteness we have worked in one construction
based on SU(5) proposed by Hall and Nomura.
In the setup the first two generations of the ten-plet matter fields
live in the bulk as the gauge and Higgs multiplets,
while the third generation ten-plet
and all of the three five-plet matters are localized on the SU(5)
conserving brane.
Given the matter geometry,
(i) realistic fermion masses and mixings are reproduced retaining
the bottom-tau mass unification, 
(ii) flavor changing processes via superparticles are suppressed due to 
the partial mass degeneracy, 
and (iii) dimension six nucleon decays mediated by the heavy gauge bosons
are avoided.
We explore observable flavor-violating signatures 
coming from the non-degenerate sfermion mass
spectra for the ten-plet sector: the left-handed squarks, the
up-type squarks and the singlet sleptons.
We have pointed out that 
CP-violating up-type squark mixings between
the first and third generations 
in both the left- and right-handed sectors
induces a sizable CEDM of the up quark, 
which is enhanced by the large top quark mass.
As a result,
the hadronic EDM constraints on the orbifold GUT framework
are predicted to be close to the current experimental bounds
as well as that from $\mu \to e \gamma$ decay.
On the contrary,
in the context of conventional four-dimensional SUSY GUTs,
the induced up quark CEDM is negligible whereas a detectable rate for 
the $\mu \to e \gamma$ decay is expected.
The future EDM and LFV experiments will act as good probes
to discriminate various SUSY GUT models.

\section*{Acknowledgments}
The work of J.H. is supported in part by the Grant-in-Aid for Science
Research, Ministry of Education, Science and Culture, Japan
(No.13135207, No.15540255 and No.17043003).  
That of M.K. is supported in part by JSPS.



\begin{thebibliography}{99}


\bibitem{Hisano:2003bd}
J.~Hisano and Y.~Shimizu,
Phys.\ Lett.\ B {\bf 565} (2003) 183.



\bibitem{Ciuchini:2003rg}
M.~Ciuchini, A.~Masiero, L.~Silvestrini, S.~K.~Vempati and O.~Vives,
Phys.\ Rev.\ Lett.\  {\bf 92} (2004) 071801.



\bibitem{Hisano:2003iw}
J.~Hisano and Y.~Shimizu,
Phys.\ Lett.\ B {\bf 581} (2004) 224.



\bibitem{Hisano:2004tf}
J.~Hisano and Y.~Shimizu,
Phys.\ Rev.\ D {\bf 70} (2004) 093001.



\bibitem{BphiK}
K.~Abe {\it et al.}  [BELLE Collaboration],
arXiv:hep-ex/0409049;
B.~Aubert {\it et al.}  [BABAR Collaboration],
arXiv:hep-ex/0408072.



\bibitem{Moroi:2000tk}
T.~Moroi,
Phys.\ Lett.\ B {\bf 493} (2000) 366.



\bibitem{Dimopoulos:1994gj}
S.~Dimopoulos and L.~J.~Hall,
Phys.\ Lett.\ B {\bf 344} (1995) 185.



\bibitem{Khriplovich:1996gk}
I.~B.~Khriplovich and K.~N.~Zyablyuk,
Phys.\ Lett.\ B {\bf 383} (1996) 429.



\bibitem{Romanino:1996cn}
A.~Romanino and A.~Strumia,
Nucl.\ Phys.\ B {\bf 490} (1997) 3.



\bibitem{Hisano:2004pw}
J.~Hisano, M.~Kakizaki, M.~Nagai and Y.~Shimizu,
Phys.\ Lett.\ B {\bf 604} (2004) 216.



\bibitem{proton_decay}
N.~Sakai and T.~Yanagida,
Nucl.\ Phys.\ B {\bf 197} (1982) 533;
S.~Weinberg,
Phys.\ Rev.\ D {\bf 26} (1982) 287.



\bibitem{dim5sup}
For examples of the four-dimensional SUSY GUTs, see following articles,
J.~Hisano, T.~Moroi, K.~Tobe and T.~Yanagida,
Phys.\ Lett.\ B {\bf 342} (1995) 138;
K.~S.~Babu and S.~M.~Barr,
Phys.\ Rev.\ D {\bf 48} (1993) 5354;
Phys.\ Rev.\ D {\bf 65} (2002) 095009;
N.~Maekawa,
Prog.\ Theor.\ Phys.\  {\bf 107} (2002) 597;
N.~Maekawa and T.~Yamashita,
Prog.\ Theor.\ Phys.\  {\bf 107} (2002) 1201;
T.~Yanagida,
Phys.\ Lett.\ B {\bf 344} (1995) 211;
T.~Hotta, K.~I.~Izawa and T.~Yanagida,
Phys.\ Rev.\ D {\bf 53} (1996) 3913;
Phys.\ Rev.\ D {\bf 54} (1996) 6970.



\bibitem{Kawamura:2000ev}
Y.~Kawamura,
Prog.\ Theor.\ Phys.\  {\bf 105} (2001) 999.



\bibitem{Altarelli:2001qj}
G.~Altarelli and F.~Feruglio,
Phys.\ Lett.\ B {\bf 511} (2001) 257.



\bibitem{Hall:2001pg}
L.~J.~Hall and Y.~Nomura,
Phys.\ Rev.\ D {\bf 64} (2001) 055003.



\bibitem{Hebecker:2001wq}
A.~Hebecker and J.~March-Russell,
Nucl.\ Phys.\ B {\bf 613} (2001) 3.



\bibitem{Hebecker:2001jb}
A.~Hebecker and J.~March-Russell,
Nucl.\ Phys.\ B {\bf 625} (2002) 128.



\bibitem{Asaka:2001eh}
T.~Asaka, W.~Buchmuller and L.~Covi,
Phys.\ Lett.\ B {\bf 523} (2001) 199.




\bibitem{Hall:2001rz}
L.~Hall, J.~March-Russell, T.~Okui and D.~R.~Smith,
JHEP {\bf 0409} (2004) 026.



\bibitem{Hall:2001xb}
L.~J.~Hall and Y.~Nomura,
Phys.\ Rev.\ D {\bf 65} (2002) 125012.



\bibitem{Hall:2002ci}
L.~J.~Hall and Y.~Nomura,
Phys.\ Rev.\ D {\bf 66} (2002) 075004.



\bibitem{Scherk-Schwarz}
J.~Scherk and J.~H.~Schwarz,
Phys.\ Lett.\ B {\bf 82} (1979) 60;
J.~Scherk and J.~H.~Schwarz,
Nucl.\ Phys.\ B {\bf 153} (1979) 61.



\bibitem{Barbieri:2001yz}
R.~Barbieri, L.~J.~Hall and Y.~Nomura,
Phys.\ Rev.\ D {\bf 66} (2002) 045025.



\bibitem{Semertzidis:2003iq}
Y.~K.~Semertzidis {\it et al.}  [EDM Collaboration],
AIP Conf.\ Proc.\  {\bf 698} (2004) 200.



\bibitem{Bigi:1991rh}
I.~I.~Y.~Bigi and N.~G.~Uraltsev,
Sov.\ Phys.\ JETP {\bf 73} (1991) 198.



\bibitem{Pospelov:2005pr}
M.~Pospelov and A.~Ritz,
arXiv:hep-ph/0504231,
and references therein.



\bibitem{Harris:1999jx}
P.~G.~Harris {\it et al.},
Phys.\ Rev.\ Lett.\  {\bf 82} (1999) 904.



\bibitem{Romalis:2000mg}
M.~V.~Romalis, W.~C.~Griffith and E.~N.~Fortson,
Phys.\ Rev.\ Lett.\  {\bf 86} (2001) 2505.



\bibitem{mass_insertion}
L.~J.~Hall, V.~A.~Kostelecky and S.~Raby,
Nucl.\ Phys.\ B {\bf 267} (1986) 415;
F.~Gabbiani, E.~Gabrielli, A.~Masiero and L.~Silvestrini,
Nucl.\ Phys.\ B {\bf 477} (1996) 321.



\bibitem{seesaw} 
M. Gell-Mann, P. Ramond and R. Slansky, Proceedings of   
the Supergravity Stony Brook Workshop, New York, 1979, eds. P. Van
Nieuwenhuizen and D. Freedman (North-Holland, Amsterdam); T. Yanagida,
Proceedings of the Workshop on Unified Theories and Baryon Number in
the Universe, Tsukuba, Japan 1979 (edited by A.  Sawada and A.
Sugamoto, KEK Report No.  79-18, Tsukuba).



\bibitem{SUSY_GUT_flavor}
R.~Barbieri and L.~J.~Hall,
Phys.\ Lett.\ B {\bf 338} (1994) 212;
R.~Barbieri, L.~J.~Hall and A.~Strumia,
Nucl.\ Phys.\ B {\bf 445} (1995) 219;
R.~Barbieri, L.~J.~Hall and A.~Strumia,
Nucl.\ Phys.\ B {\bf 449} (1995) 437.



\bibitem{Brooks:1999pu}
M.~L.~Brooks {\it et al.}  [MEGA Collaboration],
Phys.\ Rev.\ Lett.\  {\bf 83} (1999) 1521.



\bibitem{MEG}
T.~Mori, Talk given at
the 12th International Conference on Supersymmetry
and Unification of Fundamental Interactions,
SUSY 2004, 17-23 June 2004,
Tsukuba, Japan.



\bibitem{higher_dim_op}
N.~Arkani-Hamed, H.~C.~Cheng and L.~J.~Hall,
Phys.\ Rev.\ D {\bf 53} (1996) 413;
J.~Hisano, D.~Nomura, Y.~Okada, Y.~Shimizu and M.~Tanaka,
Phys.\ Rev.\ D {\bf 58} (1998) 116010.



\bibitem{Hisano:1996qq}
J.~Hisano, T.~Moroi, K.~Tobe and M.~Yamaguchi,
Phys.\ Lett.\ B {\bf 391} (1997) 341
[Erratum-ibid.\ B {\bf 397} (1997) 357].



\end{thebibliography}
\end{document}